\documentclass[titlepage,12pt]{article}
\usepackage{amssymb,psfig,epsfig,pslatex}

\usepackage[latin1]{inputenc}
\usepackage[T1]{fontenc}
\def\pb{p_b}
\def\pt{p_t}
\def\e{\epsilon}
\def\Li#1{{\mbox{Li}_#1}}
\def\mt{m_t}

\def\nn{\nonumber}
\def\pbhat{{\hat{\bf p}_b}}

\begin{document}
\begin{titlepage}
\noindent
PITHA 02/07 \hfill April 2002 \\
DESY 02-055 \hfill \, \\
TTP02-04    \hfill \, 
\vspace{0.4cm}
\renewcommand{\thefootnote}{\fnsymbol{footnote}}
\begin{center}
{\LARGE {\bf 
QCD-corrected spin analysing power of jets in decays of polarized 
top quarks}} \\
\vspace{2cm}
{\bf
A. Brandenburg $^{a,}$\footnote
{supported by a Heisenberg fellowship of D.F.G.},
Z. G. Si $^{b,}$\footnote{supported by BMBF contract 05 HT1 PAA 4}
and P. Uwer $^c$}
\par\vspace{1cm}
$^a$  DESY-Theorie, 22603 Hamburg, Germany\\
$^b$ Institut f.\ Theoretische Physik, RWTH Aachen, 52056 Aachen, Germany\\
$^c$ Institut f. Theoretische Teilchenphysik, Universit\"at Karlsruhe,
76128 Karlsruhe, Germany
\par\vspace{2cm}
{\bf Abstract:}\\
\parbox[t]{\textwidth}
{We present results for the differential distributions of 
jets from non-leptonic decays of polarized
top quarks within the Standard Model, including QCD radiative
corrections. Our work extends existing results
which are only available for semileptonic top quark 
decays at the parton level. 
For $t(\uparrow)\to b$-jet $+2$ light jets 
we compute in particular the QCD-corrected top-spin
analysing power of the $b$-quark jet and the least energetic
light jet. The dependence of the results on the choice
of the jet recombination scheme is found to be small.
In addition we compute the spin analysing
power of the thrust axis. Our results constitute a so far 
missing ingredient to analyse top quark production and 
subsequent non-leptonic decay at next-to-leading order in $\alpha_s$, keeping
the full information on the top quark polarization.}
\end{center}
\vspace{2cm}
PACS number(s): 12.38.Bx, 13.88.+e, 14.65.Ha
\end{titlepage}
\renewcommand{\thefootnote}{\arabic{footnote}}

\setcounter{footnote}{0}
\section{Introduction}
The detailed analysis of the dynamics of top quark production and decay 
is a major objective of experiments at the Tevatron, the LHC, and a possible
future linear $e^+e^-$ collider. A special feature of the top quark that
makes such studies very attractive is its large decay width: In
contrast to the light quarks the large top decay width $\Gamma_t\approx 1.5$
GeV  serves as a cut-off for non-perturbative
effects in top quark decays. As a consequence 
precise theoretical predictions of
cross sections and differential distributions involving top quarks 
are possible within the Standard Model and its extensions.
A confrontation  of such predictions 
with forthcoming high-precision data will lead
to accurate determinations of Standard Model parameters and maybe
hints to new phenomena. In particular, observables related
to the spin of the top quark can be studied and 
utilized to search for new interactions
of the top quark \cite{newphysics}.
In $e^+e^-$ collisions,
top quarks are produced highly polarized, especially  if one 
tunes the polarization of the incoming beams, as possible e.g. at
the TESLA collider \cite{TDR}. Furthermore, even for purely QCD-induced
production of top quark pairs, the spins of $t$ and $\bar{t}$ are
in general highly correlated \cite{abc}. 
\par
The polarization 
and  spin correlations of top quarks must be traced in the differential
distributions of the decay products. This is possible since the 
information on the top
polarization is transferred to the angular distribution of
the decay products through its weak,
parity violating decays. To be more precise: Consider
a polarized  ensemble of top quarks at rest with polarization
vector ${\bf P},\ 0\le |{\bf P}|\le 1$.   
The differential decay distribution with respect to
the angle $\vartheta$ between ${\bf P}$ and the direction $\hat{\bf p}$
of a given decay product reads: 
\begin{eqnarray}\label{power}
\frac{1}{\Gamma}\frac{d\Gamma}{d\cos\vartheta}=\frac{1}{2}\left(1+
|{\bf P}|\kappa_p\cos\vartheta\right).
\end{eqnarray} 
In Eq.~(\ref{power}), ${\Gamma}$ is the partial width for the 
corresponding decay
of unpolarized top quarks, and $\kappa_p$ is the so-called
{\it spin analysing power} of the final state 
particle or jet under consideration. For example, in the semileptonic decay
$t\to l^+\nu_l b$, the charged 
lepton ($b$-quark) has spin analysing power $\kappa_p=+1$ ($\sim -0.41$) 
at the tree level within the Standard Model. 
In hadronic top decays $t\to b \bar{d} u$ (where $d (u)$ stands generically
for $d,s\ (u,c)$),   
the r\^ole of the charged lepton is played by the  $\bar{d}$
quark. Only for $t\to b\bar{s}c$ the maximal spin analysing power of
the $\bar{s}$-quark
could in principle be used by tagging $b$ and $c$-quark jets. However, the
efficiency of charm-tagging is quite low, and one should try to use
a spin analyser that is both efficiently detected and has a large analysing 
power. A good choice is the least energetic light (i.e. non-$b$-quark) 
jet \cite{Jezabek:1994},
which at tree level has  $\kappa_p=+0.51$. This follows from the fact that 
with a probability of 61\% (at tree level) this jet contains the $\bar{d}$
quark.
\par
The topic of this letter are the QCD corrections
to the above tree level results for $\kappa_p$. 
In fact we will
be a little more general and 
discuss corrections to the fully differential decay 
distribution of polarized top quarks to be defined in section \ref{tree}. 
These corrections are one 
ingredient for a full analysis of
top quark (pair) production and decay at next-to-leading
order in $\alpha_s$, both at lepton
and hadron colliders. They form part of the {\it factorizable}
corrections within the pole approximation 
\cite{Stuart:1991,Aeppli:1994} for the top quark propagator(s). 
(For the non-factorizable contributions, see ref.~\cite{Beenakker:1999}.)
QCD corrections to the production of top quark pairs, including the 
full information about their spins, can be found in ref.~\cite{epem} for 
$e^+e^-$ collisions and in ref.~\cite{pp} for hadron-hadron collisons.
In the case of $e^+e^-$ collisions, also fully analytic results 
for the top quark polarization \cite{epeman1} and a specific spin correlation 
\cite{epeman2} to order $\alpha_s$ are available. 
In ref.~\cite{schmidt}, helicity amplitudes for $e^+e^-\to t\bar{t}X$ 
production
and semileptonic decays are computed to order $\alpha_s$ and are used
to construct an event generator. The reaction 
$e^+e^-\to t\bar{t}\to W^+bW^-\bar{b}$ is treated in ref.~\cite{mac:2001}, 
including both factorizable and non-factorizable QCD corrections.
The theoretical status of polarized top quark decay is as follows:
A complete calculation of the angular decay distribution for
$t(\uparrow)\to W^+ b $ to order $\alpha_s$ can be found in
ref.~\cite{Fischer:2001}.
The QCD corrections for semileptonic
polarized top quark decays have been computed
in ref.~\cite{Czarnecki:1991}. We will compare our results, from which the
semileptonic case can be easily derived, 
to  those of ref.~\cite{Czarnecki:1991}
in section \ref{vir}.
\par
The outline of this paper is as follows: In the next section we shortly
review the tree level results for the decay 
of polarized top quarks. In section
\ref{vir} we discuss the calculation of the QCD corrections. Section \ref{res}
contains our numerical results, which we discuss in section \ref{sum}. 
\section{Kinematics and tree level results}
\label{tree}
Consider an initial state consisting of top quarks
at rest with polarization ${\bf P}$. For the non-leptonic decay
\begin{eqnarray}\label{reac}
t(\pt)\to b(\pb) + u(p_u) +\bar{d}(p_{\bar{d}}),
\end{eqnarray} 
the phase space $R_3$ of the final state may be parametrized
by two scaled energies and two angles:
\begin{eqnarray}
dR_3=\frac{m_t^2}{32(2\pi)^4}dx_{\bar{d}}
dx_bd\chi d\cos\theta,
\end{eqnarray} 
where $x_b=2E_b/m_t,\ x_{\bar{d}}=2E_{\bar{d}}/m_t$, 
$\cos\theta=\hat{\bf P}\cdot
\hat{\bf{p}}_{\bar{d}}$, and $\chi$ 
is the (signed) angle
between the plane spanned by  $\hat{\bf P}$ and $\hat{\bf p}_{\bar{d}}$
and the plane spanned by $\hat{\bf p}_{\bar{d}}$ and $\pbhat$.
We will neglect the masses of the light quarks $u$ and $\bar{d}$. %
\def\pu{p_u}%
\def\pd{p_{\bar d}}%
\def\cM{{\cal M}}%
The differential decay rate is given by 
\begin{eqnarray}\label{treeres1}
d\Gamma^0 = \frac{1}{2m_t} |\cM(\pb,\pu,\pd)|^2dR_3,
\end{eqnarray}
where $|\cM(\pb,\pu,\pd)|^2$ stands for the squared matrix element
averaged over the colour of the initial state and summed over
colour and spins of the final state. 
The fully differential distribution 
for reaction~(\ref{reac}) reads at tree level:
\begin{eqnarray}\label{treeres}
\frac{d\Gamma^0}{dx_{\bar{d}}dx_bd\chi d\cos\theta}&=&
c \frac{x_{\bar{d}}(1-x_{\bar{d}}-z_b)}
{(1-x_b+z_b-\xi)^2+\eta^2\xi^2}\left(1+|{\bf P}|\cos\theta\right),
\end{eqnarray}
where   
 \begin{eqnarray}\label{kappa}
 c=N_C|V_{ud}|^2\frac{e^4|V_{tb}|^2m_t}{128(2\pi)^4\sin^4\theta_W
}= N_C|V_{ud}|^2\frac{|V_{tb}|^2m_tG_F^2m_W^4}{4(2\pi)^4}, 
\end{eqnarray}
with
 \begin{eqnarray}
\xi = \frac{m_W^2}{m_t^2},\ \ \ \ \ \eta = \frac{\Gamma_W}{m_W},\ \ \ \ \ \
z_b=\frac{m_b^2}{m_t^2}.
\end{eqnarray}
A convenient way to compute the spin analysing power $\kappa_p$ 
defined in Eq.~(\ref{power}) is to evaluate the expectation value
of $\cos\vartheta$:
\begin{eqnarray}
\langle\cos\vartheta\rangle \equiv \langle\hat{{\bf p}}\cdot \hat{\bf P} \rangle=\frac{1}{\Gamma}\int d\Gamma \cos\vartheta
=\frac{\kappa_p|{\bf P}|}{3}.
\end{eqnarray} 
We want to compute $\kappa_p$ 
for the following
choices of $\hat{\bf p}$:
\begin{eqnarray}
 ({\rm i})&&\ \ \ \hat{\bf p}= \hat{\bf p}_{\bar{d}}, \nonumber  \\
 ({\rm ii})&&\ \ \ \hat{\bf p}= \pbhat, \nonumber \\
 ({\rm iii})&&\ \ \ \hat{\bf p}= \hat{\bf p}_u \nonumber  \\
 ({\rm iv})&&\ \ \ \hat{\bf p}= \hat{\bf k}_j,
 \nonumber  \\
 ({\rm v})&&\ \ \ \hat{\bf p}= {\bf T}.
 \label{eq:obsdef}
\end{eqnarray}
In $({\rm iv})$, $\hat{\bf k}_j$ is the direction of the
light (non-$b$-quark) jet with the {\it  
smallest} energy. In the leading order calculation of $\kappa_p$, 
one can simply identify jets with the partons, and thus
$\hat{\bf k}_j$ denotes the direction of the up-type quark if
$E_u<E_{\bar{d}}$, and the direction of the down-type
quark otherwise.
Finally, in $({\rm v})$, ${\bf T}$ denotes the 
{\it thrust axis} \cite{Fa77}. We define the orientation of the 
thrust axis such that ${\bf T}\cdot 
\pbhat$ is positive. In leading order, the oriented thrust is given
by $\hat{\bf a}\ {\rm sign}\left(\hat{\bf a}\cdot\pbhat\right)$, where
$\hat{\bf a}$ denotes the direction of the parton with the largest 3-momentum.
\par
In Table 1 we list our
results. As input we use  $m_t = 175$ GeV, $m_b = 5$ GeV, 
$m_W = 80.41$ GeV, and  
$\Gamma_W = 2.06$ GeV. All other constants cancel in the computation
of $\kappa_p$.
\begin{table}[h]
\caption{Born results for spin analysing power of $\bar{d}$, $b$, $u$ 
, least energetic light jet and thrust axis.}
\begin{center}
\begin{tabular}{|c|c|c|c|c|}\hline
& $m_b=0$, $\Gamma_W\to 0$ &$m_b=0$, $\Gamma_W$ kept  & $m_b=5$ GeV,
 $\Gamma_W\to 0$ & $m_b=5$ GeV,
 $\Gamma_W$ kept\\  \hline
$\kappa_{\bar{d}}^0$ & $1$&  $1$ & 
$1$ & $1$\\
$\kappa_{b}^0$ & $-0.40622
$& $-0.40867$ & 
$ -0.40553$  & 
$-0.40800$\\
$\kappa_{u}^0$ & $-0.31817$&  $-0.31091$ & 
$-0.31964$ & $-0.31236$\\
$\kappa_j^0$ & $\ \ \ 0.50774$ & $\ \ \  0.51088$ & 
$\ \ \ 0.50708$ & $\ \ \ 0.51021$\\ 
$\kappa_T^0$ & $-0.31712$ & $-0.31782$ & 
$-0.31597$ & $-0.31671$\\\hline
\end{tabular}
\vspace*{1em}
\label{tab:lo}
\end{center}
\end{table}
We also give numbers for the limiting cases
$m_b=0$ and $\Gamma_W\to 0$.
The latter corresponds to the {\it narrow width approximation} for the 
$W$-boson, i.e. the replacement
\begin{eqnarray}
\frac{1}{(k^2-m_W^2)^2+m_W^2\Gamma_W^2}\to \frac{\pi}{m_W\Gamma_W}
\delta(k^2-m_W^2),
\end{eqnarray}
where $k^2$ is the squared momentum of the $W$.
One sees that the results are essentially
insensitive to the bottom quark mass, while keeping the  $W$ width
changes, e.g., $\kappa_u$ by more than 2\% as compared to the 
narrow width approximation. One further comment concerning the dependence
of $\kappa_p$ on the $W$ boson width: In the narrow width approximation,
$\Gamma_W$ drops out completely. (This is also true for the QCD-corrected
results for $\kappa_p$.) Keeping the Breit-Wigner form for the $W$ propagator,
the dependence of $\kappa_p$ on the $W$ width is extremely small.
For example, 
a 10\% change in $\Gamma_W$ 
changes the top quark hadronic decay rate $\Gamma$ by about 10\%, but the
spin analysing powers are affected only at the permill level.
Therefore we simply use, both at leading and next-to-leading order,
a fixed value $\Gamma_W=2.06$ GeV.

As already mentioned in the introduction, 
the case of semileptonic decays $t\to l^+\nu_l b$ follows
from the hadronic decay by the identifications
$l^+ \leftrightarrow \bar{d},\ \nu_l \leftrightarrow u$, and
by leaving out the factor $N_C|V_{ud}|^2$ in Eq.~(\ref{kappa}). 

\section{QCD corrections}\label{vir}
The computation of the QCD corrections to the decay
$t(\uparrow)\to b \bar{d}u$
is a generalization
of the corresponding computation 
for $t(\uparrow)\to b l^+\nu_{l}$ \cite{Czarnecki:1991}:
For the virtual amplitude, one has to add the $W \bar{d}u$
gluonic vertex correction (box diagrams do not contribute). The
emission of real gluons from $u$ and $\bar{d}$ does not interfere
with the gluon emission from $t$ and $b$ and are added
incoherently. 
\par
We work in $d=4-2\epsilon$ space-time dimensions to regularize
both soft/collinear and ultraviolet singularities. 
We simplify the virtual amplitude using an anticommuting
$\gamma_5$, thus respecting the chiral Ward identities.
The only divergent part
of the amplitude is proportional to the Born amplitude.
The square of the Born amplitude does not depend on $d$ if
one keeps the $W$-boson polarization in 4 dimensions. Therefore,
we can evaluate all necessary traces in 4 dimensions.
We also choose to keep the phase space measure $dR_3$ in 4 dimensions.
Our result for the virtual corrections reads:
\begin{eqnarray}\label{virres}
\frac{d\Gamma^{\rm virtual}}{dx_{\bar{d}}dx_bd\chi d\cos\theta}&=&
\frac{c}{(1-x_b+z_b-\xi)^2+\eta^2\xi^2}\frac{\alpha_sC_F}{4\pi}
\left(\frac{4\pi\mu^2}{m_t^2}\right)^{\epsilon}
\frac{1}{\Gamma(1-\epsilon)}\nonumber \\ &\times&
\left[f_1(x_{\bar{d}},x_b,z_b)\left(1+|{\bf P}|\cos\theta\right)
+f_2(x_{\bar{d}},x_b,z_b)|{\bf P}|\sin\theta\cos\chi\right],
\end{eqnarray}
with
\begin{eqnarray}\label{f1}
f_1(x_{\bar{d}},x_b,z_b)&=& \Bigg\{
\frac{1}{\beta}\Bigg[
-2\left(1+\frac{1}{\epsilon}\right)\ln(\omega)+\ln^2(\omega)
+4\ln(\omega)\ln\left(\frac{x_b(1-\omega)}{1+\omega-x_b\omega}\right)
\nonumber \\
&+& 4{\rm Li}_2\left(\frac{\omega(1+\omega-x_b)}{1+\omega-x_b\omega}\right)
-  4{\rm Li}_2\left(\frac{1+\omega-x_b}{1+\omega-x_b\omega}\right)
\Bigg]\nonumber \\
&-&\frac{4}{\epsilon}-8+\ln(z_b)
+g^{Wu\bar{d}}\Bigg\}x_{\bar{d}}(1-x_{\bar{d}}-z_b)\nonumber \\
&+& \left[\frac{\ln(\omega)}{\beta}-\ln(z_b)\right]
\left[(1-x_{\bar{d}}-z_b)^2+z_b(1-z_b)\right]\nonumber \\
&+&\frac{2z_b\ln(\omega)}{x_b\beta}(2z_b-x_b+2x_{\bar{d}}-x_{\bar{d}}x_b)
+O(\epsilon),\label{eq:res1}
\end{eqnarray}
\begin{eqnarray}\label{f2}
f_2(x_{\bar{d}},x_b,z_b)&=& \frac{\beta\sin\theta_{\bar{d}b}
x_{\bar{d}}(1-x_{\bar{d}}-z_b)}{2(1-x_b+z_b)}
\Bigg\{x_b\left[\frac{\ln(\omega)}{\beta}-\ln(z_b)\right]\nonumber \\
&+&2z_b
\left[\frac{1-x_b+z_b}{1-x_{\bar{d}}-z_b}-1\right]
\frac{\ln(\omega)}{\beta}\Bigg\}+O(\epsilon),
\end{eqnarray}
where
\begin{eqnarray}
\beta=\sqrt{1-4z_b/x_b^2},
\end{eqnarray}
\begin{eqnarray}
\omega=\frac{1-\beta}{1+\beta},
\end{eqnarray}
\begin{eqnarray}\label{Wud}
g^{Wu\bar{d}}= 
2\left(\frac{m_t^2}{k^2}\right)^\epsilon\left[-\frac{2}{\epsilon^2}
-\frac{3}{\epsilon}-8+\pi^2\right],
\end{eqnarray}
and $\theta_{\bar{d}b}$ is the angle between the  $\bar{d}$ and the $b$
in the top quark rest frame.
One can easily recover the case of semileptonic decays from the above result
by leaving out a factor $N_C|V_{ud}|^2$ and 
the additional incoherent contribution (\ref{Wud}) from the 
QCD-corrected $W u \bar{d}$ vertex.
Thus we can perform an analytic comparison to the results given in Eqs.
(2.7), (2.8) of  ref.~\cite{Czarnecki:1991}. We  find complete agreement
by using the well-known correspondence 
\begin{eqnarray}
\frac{1}{\epsilon}
\left(\frac{4\pi\mu^2}{m_t^2}\right)^\epsilon
\frac{1}{\Gamma(1-\epsilon)}\rightarrow
\ln\left(\frac{\lambda^2}{m_t^2}\right)
\end{eqnarray} 
between dimensional regularization 
in $d=4-2\epsilon$ dimensions at a scale $\mu$
and regularization by a small gluon mass $\lambda$ used in ref.~\cite{Czarnecki:1991}.
\par
In addition to the virtual corrections we have to include the real
corrections which are given by the process with an additional gluon:
\def\pg{p_g}
\begin{equation}
  t(\pt)\to b(\pb) + u(p_u) +\bar{d}(p_{\bar{d}}) + g(\pg).
\end{equation}
This
contribution can also be split into two separate parts, which do
not interfere with each other: the case where
the gluon is emitted from a  heavy quark ($t$ or $b$), 
and the contribution
where the gluon is emitted from a light quark. The calculation 
of the corresponding amplitudes is straightforward. As always in the  
case of a jet-calculation we have to address the question how to cancel
the infrared and collinear singularities. For the contribution where
the gluon is emitted from the secondary fermion line we use the dipole 
formalism \cite{Catani:1996vz}. Dropping overall factors the subtraction term
for this process is given by
\begin{equation}
 - {{V}_{ug,\bar d}\over 2(\pu\cdot\pg)}
  |\cM(\pb,\tilde p_{ug},\pd)|^2
  - {{V}_{\bar dg,u}\over 2(\pd\cdot\pg)}
  |\cM(\pb,\pu,\tilde p_{{\bar d }g})|^2
  \label{eq:subterm}
\end{equation}
with $\cM(\pb,\pu,\pd)$ being the matrix element at leading order and 
$\tilde p $, $V$  are defined in Eq.~(5.3) and Eq.~(5.7) of ref. 
\cite{Catani:1996vz}. 
The subtraction term integrated over the singular region 
is also given in ref. \cite{Catani:1996vz} and reads:
\begin{equation}
\frac{\alpha_sC_F}{2\pi}\left(\frac{4\pi\mu^2}{k^2}\right)^{\epsilon}
\frac{1}{\Gamma(1-\epsilon)}
\left\{\frac{2}{\epsilon^2}+\frac{3}{\epsilon}+10-\pi^2+O(\epsilon)\right\}
|\cM(\pb,\pu,\pd)|^2.
\end{equation}
Comparing with Eq.~(\ref{Wud}) one obtains immediately the cancellation
of the soft and collinear singularities. For the case where the 
gluon is emitted from the heavy quark line only a soft singularity
is present. The collinear singularity is regulated by the finite quark masses.
To extract the soft singularity we slice the phase space as follows:
\def\xmin{{x_{\mbox{\scriptsize min}}}}
\begin{equation}
   1 = \Theta(2 (\pt\cdot\pg) - \xmin m_b\*m_t) 
    + \Theta(\xmin m_b m_t- 2 (\pt\cdot\pg) ).
    \label{eq:slicing}
\end{equation}
This splits the phase space into a `resolved' and an `unresolved' region.
The contribution from the resolved region is obtained from a numerical
integration in 4 dimensions. In the unresolved region one can use the
soft factorization to approximate the matrix element and  integrate 
out the soft gluon. The result is
given by:
\begin{eqnarray}
  &&{1\over 2m_t}\int dR_4^d |\cM(\pb,\pu,\pd,\pg)|^2 
  \Theta(\xmin m_b m_t- 2 (\pt\cdot\pg) )\nn\\
  &&={\alpha_s C_F\over 2\pi} 
  \left({{4\pi\mu^2\over x_{\rm min}^2\mt^2}}\right)^{\e} 
\*{1\over \Gamma(1-\e)} {1\over 2m_t}
  \bigg\{
  \left[{1\over \e} -\ln(z_b)\right]
\*\left[2+{1\over \beta}\*\ln(\omega) \right]\nn\\\nn\\
  &&
  -{1 \over \beta} \*(-2\*\beta
  +2\*\Li2(1-\omega)
  +\ln(\omega)
  +{1\over 2}\*\ln(\omega)^2
  )\bigg)|\cM(\pb,\pu,\pd)|^2 dR_3^d.
\end{eqnarray}
In order to be consistent with our definition of the phase space measure
$dR_3$ for the virtual corrections, we have to evaluate the above
formula for $dR_3^{d=4}$.
Comparing with Eq.~(\ref{eq:res1}) it is straightforward to see that
the infrared singularities cancel. The numerical implementation of the
subtraction term given in Eq.~(\ref{eq:subterm}) and the slicing given in
Eq.~(\ref{eq:slicing}) does not impose any problem. One should keep in mind
that the method presented above allows the calculation of arbitrary 
infrared save observables.

\section{Numerical results}\label{res}
While for the three parton final state  
one can simply identify the partons 
with the jets, 
in the case of real gluon
emission a definition of jets in terms of partons is needed which
fulfills the condition of infrared-safeness. There are many 
variations of such definitions, and in general the NLO result for $\kappa_p$
will depend on the chosen definition. In order to study this dependence,
we choose two different jet clustering schemes: The E-algorithm and the 
Durham-algorithm. The first step is to compute for all pairs $(i,j)$ of the
momenta of the 
final state partons the jet measure $y_{ij}$, which reads
\begin{eqnarray}
y_{ij}&=& \frac{(p_i+p_j)^2}{m_t^2}\ \ \ \ \ \mbox{E-algorithm}, \\
y_{ij}&=& \frac{2{\rm min}\{E_i^2,E_j^2\}}{m_t^2}(1-\cos\theta_{ij})
\ \ \ \ \ \mbox{Durham-algorithm},
\end{eqnarray}
where $\theta_{ij}$ is the angle between parton $i$ and parton $j$ 
in the top quark rest frame. The second step is to recombine
the two partons with the smallest $y_{ij}$ into a pseudoparticle
with momentum $p_k=p_i+p_j$. Then the direction of the $b$-jet 
and the light quark jet with the smallest energy can be
readily obtained from $p_k$ and the remaining two momenta.
For $b\bar{s}cg$ events also the direction of the charm jet
can be defined if the charm is tagged.
The construction of the ``$\bar{s}$-jet'' from $b\bar{s}cg$ events
is not straightforward: Only events where both a $b$-jet and a $c$-jet
are tagged can be used (in particular, the (rare) events 
where $b$ and $c$ are clustered into a single jet have to be discarded), and
for those events the remaining third jet is defined to be the 
``$\bar{s}$-jet''. This definition includes cases where the $\bar{s}$ is
recombined with the $b$- or $c$-quark and the remaining third jet
consists of a hard gluon rather than an $\bar{s}$-quark. 
\par Note that no jet resolution parameter enters in the above definitions; 
this is not necessary, since the leading order
process is free from soft and collinear singularities.
\par
In an experiment which produces top quarks that decay into jets, the first
step in the analysis is to identify the signal by using a jet finding
 algorithm and by applying a number of cuts. For example, at the Tevatron a
cone jet algorithm is used to classify the events according to the 
number of jets. To be more specific, consider top quark pairs where the $t$
decays semileptonically and the $\bar{t}$ decays into jets. Then the event
contains at least two $b$-jets and two light jets. For those events
with 4 or more jets originating from the $\bar{t}$, our algorithm should be 
used in addition to the production-specific jet algorithm, thus leaving
only events with exactly 3 jets from the $\bar{t}$ decay.
In principle the value for $\kappa_p$ can depend on the details
of the ``pre-clustering''. This should be studied in Monte-Carlo 
simulations.
\par
We can also compute $\kappa_p$ for 
bare $b,\bar{d}\ (\bar{s})$ and $u\ (c)$ 
quarks, since the directions of a quark and a quark plus
a collinear or soft gluon are identical and thus the condition
of infrared/collinear safeness is fulfilled. This serves as a benchmark
for the realistic case of $\kappa_p$ for jets. Note however that 
recombination is needed to define
the direction of the least energetic light jet.  
In the case 
of the thrust axis the concept of a jet is not needed.
\par
We write our results to order $\alpha_s$ in the following form
\begin{eqnarray}\label{kappa_nlo}
\kappa_p = \frac{\Gamma^0\kappa_p^{0}+\alpha_s\Gamma^1\kappa_p^{1}}
{\Gamma^0+\alpha_s\Gamma^1}=\kappa_p^0+\alpha_s\frac{\Gamma^1}{\Gamma^0}\left(\kappa_p^1-\kappa_p^0\right)+O(\alpha_s^2)\equiv \kappa_p^0[1+\delta_p^{QCD}]
+O(\alpha_s^2) ,
\end{eqnarray}
where $\kappa_p^{0}$ denotes the Born result.
Table 2 gives our results for $\kappa_p$ and $\delta_p^{QCD}$, where we use
the expanded form of Eq.~(\ref{kappa_nlo}). The strong coupling 
constant is set to $\alpha_s(m_t)=0.108$. 
\begin{table}[h]
\caption{QCD-corrected results for spin analysing powers.}
\begin{center}
\begin{tabular}{
|c|c|c|c|}\hline
& partons & jets, E-alg.  & jets, D-alg. \\  \hline
$\kappa_{\bar{d}}$ 
& $0.9664(7)$&  $0.9379(8)$ & 
$0.9327(8)$ \\ 
$\delta^{QCD}_{\bar{d}}$ [\%] & $-3.36\pm 0.07$ & $-6.21\pm 0.08$
 &  $-6.73 \pm 0.08$   \\ \hline  
$\kappa_{b}$ 
& $-0.3925(6)$& $-0.3907(6)$ & 
$-0.3910(6)$\\
$\delta^{QCD}_{b}$ [\%] & $-3.80\pm 0.15$ & $-4.24\pm  0.15$ & 
$-4.18  \pm 0.15$ \\ \hline  
$\kappa_{u}$ 
& $-0.3167(6)$&  $-0.3032(6)$ & 
$-0.3054(6)$ \\
$\delta^{QCD}_{u}$ [\%] & $+1.39\pm 0.19$ & $-2.93\pm 0.19$ & 
$-2.22\pm  0.19$   \\ \hline
$\kappa_j$ 
& $-$ & $0.4736(7)$ & 
$ 0.4734(7)$\\
 $\delta^{QCD}_{j}$ [\%] & $-$ & $-7.18\pm  0.13$ &  $-7.21\pm 
 0.13$   \\ \hline
$\kappa_T$ 
& $ -0.3083(6)$ & $-$ & $-$\\
$\delta^{QCD}_{T}$ [\%] & $-2.65\pm 0.19$ & $-$ &  $-$  \\
\hline
\end{tabular}
\vspace*{1em}
\label{tab:nlo}
\end{center}
\end{table}    
\section{Discussion and conclusions}\label{sum}
Our results listed in Table 2 show that the top-spin analysing powers
of the final states in non-leptonic top quark decays receive
QCD corrections in the range $+1.4$\% to $-7.2$\%. This has to be contrasted
with the spin analysing power of the charged lepton in  
decays $t(\uparrow)\to
bl^+\nu_l$: the QCD corrected result (for $m_b=0$) 
reads \cite{Czarnecki:1991} $\kappa_l=1-0.015\alpha_s$, i.e. the correction
is at the permill level. 
The QCD corrections to the spin analysing power
of a given quark are much larger
due to hard gluon emission from that quark.
The spin analysing power of jets is smaller than that of
the corresponding bare quarks.
This effect is largest for the ``$\bar{s}$-jet''. 
We find only a small (at most $0.7$\%) 
dependence of the results on the jet algorithm.
In practice the most important spin analysers are, as far as non-leptonic 
top decays are concerned, the $b$-quark jet and
the least energetic light (non-$b$-quark) jet. The QCD corrected results are
$\kappa_b\approx -0.39$ and $\kappa_j\approx 0.47$. For the $b$-jet 
the difference between the parton level result and the jet result is small. 
The oriented thrust axis, for
which $\kappa_T\approx -0.31$, 
may also serve as a good spin analyser, since it is easily measurable.
\par
In summary, we have computed the
QCD corrections to the top-spin analysing power of jets
and the thrust axis in non-leptonic polarized top quark decays.
Our results can be used in conjunction with the known 
NLO QCD results for the production of polarized top quarks both at lepton and 
hadron colliders.  
 
\section*{Acknowledgements}
We would like to thank W. Bernreuther for suggesting this work
and him and S. Dittmaier for useful discussions.
%

\noindent

\begin{thebibliography}{99}
\bibitem{newphysics}
D.~Atwood, A.~Aeppli and A.~Soni,
Phys.\ Rev.\ Lett.\  {\bf 69}, 2754 (1992);
W.~Bernreuther, O.~Nachtmann, P.~Overmann and T.~Schr\"oder, 
Nucl.\ Phys.\ B {\bf 388} (1992) 53, Erratum-ibid. B {\bf 406} (1993) 516; 
G.~L.~Kane, G.~A.~Ladinsky and C.~P.~Yuan,
Phys.\ Rev.\ D {\bf 45}, 124 (1992);
C.~R.~Schmidt and M.~E.~Peskin,
Phys.\ Rev.\ Lett.\  {\bf 69}, 410 (1992);
W.~Bernreuther and A.~Brandenburg,
Phys.\ Lett.\ B {\bf 314}, 104 (1993);
W.~Bernreuther, A.~Brandenburg and M.~Flesch,
hep-ph/9812387;
A.~Brandenburg and J.~P.~Ma,
Phys.\ Lett.\ B {\bf 298}, 211 (1993);
P.~Haberl, O.~Nachtmann and A.~Wilch,
Phys.\ Rev.\ D {\bf 53}, 4875 (1996);
K.~Cheung,
Phys.\ Rev.\ D {\bf 55}, 4430 (1997);
B.~Grzadkowski, B.~Lampe and K.~J.~Abraham,
Phys.\ Lett.\ B {\bf 415}, 193 (1997);
W.~Bernreuther, M.~Flesch and P.~Haberl,
Phys.\ Rev.\ D {\bf 58}, 114031 (1998);
D.~Atwood, S.~Bar-Shalom, G.~Eilam and A.~Soni,
Phys.\ Rept.\  {\bf 347}, 1 (2001).
%
\bibitem{TDR}J.A. Aguilar-Saavedra et al., 
TESLA technical design report part III: Physics at an $e^+e^-$ linear
collider [hep-ph/0106315].
%
\bibitem{abc}
V.~Barger, J.~Ohnemus and R.~J.~Phillips,
Int.\ J.\ Mod.\ Phys.\  A\ 4, 617 (1989);
T.~Stelzer and S.~Willenbrock,
Phys.\ Lett.\   B\ {\bf 374}, 169 (1996);
G.~Mahlon and S.~Parke,
Phys.\ Rev.\   D\ {\bf 53}, 4886 (1996);
S. Parke and Y. Shadmi, Phys.\ Lett. B {\bf 387} (1996) 199;
A.~Brandenburg,
Phys.\ Lett.\   B\ {\bf 388}, 626 (1996);
D.~Chang, S.~Lee and A.~Sumarokov,
Phys.\ Rev.\ Lett.\  {\bf 77}, 1218 (1996);
G.~Mahlon and S.~Parke,
Phys.\ Lett.\  B\ {\bf 411}, 173 (1997).
%
\bibitem{Jezabek:1994}
M.~Jezabek, Nucl. Phys. Proc. Suppl. 37B (1994) 197, 
[hep-ph/9406411].
 
\bibitem{Stuart:1991}
R. G. Stuart, Phys. Lett. B 262 (1991) 113.

\bibitem{Aeppli:1994}
A. Aeppli, G. J. van Oldenborgh and D. Wyler, Nucl. Phys. B 428 (1994) 126
[hep-ph/9312212].

\bibitem{Beenakker:1999}
W.~Beenakker, F. A. Berends and A. P. Chapovsky,
Phys.\ Lett.\ B\ 454 (1999) 129 [hep-ph/9902304].

\bibitem{epem} 
A. Brandenburg, M. Flesch, and  P. Uwer,
Phys. Rev. D {\bf 59} (1999) 014001 [hep-ph/9806306]; 
Chechoslovak Journal of Physics, v. 50 (2000) Suppl. S1, 51-58
[hep-ph/9911249]. 
\bibitem{pp} W.~Bernreuther, A.~Brandenburg and Z.~G.~Si,
Phys.\ Lett.\ B\ 483 (2000) 99 [hep-ph/0004184];  
W. Bernreuther, A. Brandenburg, Z.G. Si and P. Uwer,
Phys. Lett. B {\bf 509} (2001), 53 [hep-ph/0104096];
Phys. Rev. Lett. 87 (2001) 242002 [hep-ph/0107086].
\bibitem{epeman1} 
J.G. K\"orner, A. Pilaftsis, and M.M. Tung, Z. Phys. C {\bf 63} 
(1994) 575; M.M. Tung, Phys. Rev. D {\bf 52} (1995) 1353;
S. Groote and J.G. K\"orner,  
Z. Phys. C {\bf 72} (1996) 255; 
V. Ravindran, W.L. van Neerven, Nucl. Phys. B589 (2000) 507 
[hep-ph/0006125] 
\bibitem{epeman2} M.M. Tung,  J. Bernabeu,  and J. Penarrocha,
Phys. Lett. B {\bf 418} (1998) 181; S. Groote, J.G. K\"orner, and J.A. Leyva,
Phys. Lett. B {\bf 418} (1998) 192.
\bibitem{schmidt} C. Schmidt, Phys. Rev. D {\bf 54} (1996) 3250.  
\bibitem{mac:2001} C. Macesanu, UR-1646, OSU-HEP-01-13, hep-ph/0112142. 

\bibitem{Fischer:2001}
M.~Fischer, S.~Groote, J.~G.~K\"orner, M.~C.~Mauser,
Phys. Rev. D\ 65 (2002) 054036,  [hep-ph/0101322].

\bibitem{Czarnecki:1991}
M. Jezabek and J. H. K\"uhn, Nucl. Phys. B {\bf 320}, 20 (1989);
A.~Czarnecki, M.~Jezabek and J.~H.~K\"uhn,
Nucl.\ Phys.\ B {\bf 351}, 70 (1991).




\bibitem{Catani:1996vz}
S.~Catani and M.~H.~Seymour,
Nucl.\ Phys.\ B {\bf 485} (1997) 291
[Erratum-ibid.\ B {\bf 510} (1997) 503]

\bibitem{Fa77}
E. Farhi, Phys. Rev. Lett. {\bf 39} (1977) 1587.




\end{thebibliography}
\end{document}